\documentclass[reprint,superscriptaddress,amsmath,amssymb,aps]{revtex4-2}

\usepackage[utf8]{inputenc}
\usepackage{graphicx}
\usepackage{color}
\usepackage{physics}
\usepackage{braket}
\usepackage{comment}
\usepackage{mathrsfs}
\usepackage{enumerate}

\usepackage[version=4]{mhchem}
\usepackage[colorlinks=true,
            linkcolor=blue,
            citecolor=blue,
            urlcolor=blue]{hyperref}
\usepackage{lipsum}
\usepackage{algorithmic}
\usepackage{algorithm}
\usepackage[whole]{bxcjkjatype}

\date{\today}
\usepackage{amsthm}

\begin{document}
\title{Efficient Quantum Algorithm for Solving Linear Distributed Delay Differential Equations}
\author{Wataru Setoyama}
\email{wataru.setoyama@riken.jp}
\affiliation{RIKEN Center for Quantum Computing (RQC), Hirosawa 2-1, Wako, Saitama 351-0198, Japan}

\author{Keisuke Fujii}
\affiliation{RIKEN Center for Quantum Computing (RQC), Hirosawa 2-1, Wako, Saitama 351-0198, Japan}
\affiliation{Graduate School of Engineering Science, Osaka University,
1-3 Machikaneyama, Toyonaka, Osaka 560-8531, Japan.}
\affiliation{Center for Quantum Information and Quantum Biology, Osaka University, 560-0043, Japan.}

\date{\today}

\begin{abstract}
\noindent Non-Markovian dynamics is ubiquitous in both quantum and classical systems, but the numerical computation of the time-delay dynamics is demanding.
In this work, we propose an efficient quantum algorithm for solving linear distributed delay differential equations and identify the condition under which it applies.
Using the linear chain trick, the distributed delay differential equations can be embedded into ordinary differential equations augmented with auxiliary variables, when the kernel function is characterized by a phase‑type distribution.
Employing the Schr\"{o}dingerization method, the resulting equations can be embedded into the Schr\"{o}dinger equation and efficiently solved by Hamiltonian simulation.
Although this embedding requires the augmented differential equation to be semi-stable, we show that it is satisfied if and only if the original distributed-delay differential equations are semi-stable.
The query complexity to obtain the normalized solution state of the $N$-dimensional delay system $\ket{\mathbf{x}(t)}\equiv\mathbf{x(t)}/\|\mathbf{x}(t)\|$ is $\mathcal{O}((st\|H\|_{\max}+\log\epsilon^{-1}/\log\log\epsilon^{-1})\|\mathbf{x}(0)\|/\|\mathbf{x}(t)\|)$ with $\epsilon$, $g$, $H$, and $s$ being the allowable error, the dimension of the auxiliary variables associated with each kernel function, the Hamiltonian operator, and its sparsity, respectively.
The gate complexity is given by this quantity multiplied by $\mathcal{O}(m+\log(N(1+gs)))$, where $m$ is the number of precision bits. 
To demonstrate the efficacy of our method, we present its applications to the generalized master equation and to the Redfield equation of the dephasing model.

\end{abstract}

\maketitle
\section{INTRODUCTION}
Non-Markovian dynamics, in which memory of past states influences temporal evolution, constitutes a subject of broad and enduring interest in both classical and quantum systems~\cite{gardiner2004quantum,kubo2012statistical}.
This class of dynamics is exemplified by phenomena such as particle transport in open systems~\cite{cortes1985generalized,darve2009computing}, epidemic outbreaks~\cite{lloyd2001realistic,kiss2017mathematics}, population dynamics with maturation~\cite{gopalsamy2013stability}, and feedback control in robotic systems~\cite{hu2003dynamics,7954629}.
These phenomena are mathematically captured by distributed delay differential equations, characterized by the presence of a convolution integral that weights past states via kernel functions~\cite{smith2010introduction,gopalsamy2013stability}.
Although this class of differential equations exhibits wide applicability, the solution of these systems entails significant mathematical and computational complexity~\cite{smith2010introduction,glass2021nonlinear}, mainly due to the requirement of the entire history of the system and the evaluation of the convolution integrals at each time step.

Quantum algorithms have recently emerged as promising methods for the efficient solution of differential equations governing classical dynamics~\cite{babbush2023exponential,berry2017quantum,liu2021efficient,engel2021linear,fang2023time,tanaka2023polynomial,krovi2023improved}.
A salient feature is the reduction in computational complexity: whereas classical algorithms require polynomial resources in the number of degrees of freedom, quantum parallelism enables solutions of high-dimensional linear differential equations with only a logarithmic number of quantum gates.
To solve linear and nonlinear differential equations, these algorithms mainly employ the Hamiltonian simulation method~\cite{berry2015hamiltonian,low2017optimal} or the Harrow-Hassidim-Lloyd algorithm~\cite{harrow2009quantum}.
However, because the former reformulates the problem as a Schr\"{o}dinger-type equation and the latter reduces it to a system of linear equations arising from stepwise time evolution, neither approach can be straightforwardly applied to differential equations that involve convolution integrals over past states.
Therefore, their applications have so far been confined to ordinary differential equations that depend only on the present state, leaving distributed delay differential equations beyond reach. 

In this work, we propose a quantum algorithm to efficiently solve linear distributed-delay differential equations by embedding these equations in the Schr\"{o}dinger equation.
Using the linear chain trick~\cite{hurtado2019generalizations}, we can convert these delay differential equations into larger-dimensional and linear ordinary differential equations when their non-Markovian effects are characterized by phase-type distributions (i.e., linear combinations or convolutions of exponential distributions).
Employing the Schr\"{o}dingerization technique~\cite{jin2024quantum,jin2024bquantum,jin2025schrodingerization}, we can encode the resulting equations into the Schr\"{o}dinger equations and efficiently solve them by Hamiltonian simulation.
However, because the Schr\"{o}dingerization method requires the semi-stability condition of the differential equations obtained from the linear chain trick, these two methods cannot be combined in a straightforward manner.
In this study, we identify the condition under which the linear distributed-delay differential equations can be mapped to the Schr\"{o}dinger equations.
Notably, when the kernel functions are of phase type, this semi-stability requirement is satisfied if and only if the linear distributed-delay differential equations are also semi-stable.
In systems where the norm represents a conserved quantity, such as total population or total probability, delay differential equations naturally exhibit semi-stable dynamics~\cite{fedotov2002non,allegrini2003generalized,breuer2002theory}. 
The proposed algorithm is particularly well suited for analyzing such systems.
 
For $N$-dimensional linear distributed-delay differential equations, our proposed algorithm can prepare the solution quantum state $\ket{\mathbf{x}(t)}=\mathbf{x}(t)/\|\mathbf{x}(t)\|$ with $\mathcal{O}\left((\|\mathbf{x}(0)\|/\|\mathbf{x}(t)\|)(\epsilon^{-1}\,s\,t\,\|\mathbf{C}\|_{\max}+\log \epsilon^{-1}/\log\log\epsilon^{-1})\right)$ queries, where $\mathbf{C}$ is the linear operator of the augmented system, $s$ is its sparsity, and $\epsilon$ is an allowable error.
The gate complexity is obtained by multiplying the query complexity by  $\mathcal{O}\left(( m + \log (N(1+gs))\right)$, where $m$ denotes the number of precision bits and $g$ the dimension of the generative matrix of the phase-type distribution.
Moreover, although the proposed method proceeds through the ordinary differential equation of the augmented system, this augmented state can be efficiently prepared provided that the initial state $\ket{\mathbf{x}(0)}$ of the original system can be efficiently prepared.
To demonstrate the effectiveness of the proposed method, we examined its applicability to the generalized master equation as a representative classical system and to the Redfield equation of the dephasing model as a representative quantum system.
Our quantum algorithm efficiently solves distributed delay differential equations and opens a new avenue for simulating large-scale non-Markovian dynamics, spanning from classical to quantum regimes.

\section{PRELIMINARY}
\subsection{Linear chain trick}
Let us consider the following distributed delay differential equations:
\begin{align}
\label{eq:simple_DDE}
    \frac{dx(t)}{dt}=ax(t)+\int_0^tK(\tau)x(t-\tau)d\tau,
\end{align}
where $a$ is a constant and $K(\tau)\geq 0$ is a kernel function which weights the past states $x(t-\tau)$ with respect to time delay $\tau$.
In the r.h.s. of Eq.~\eqref{eq:simple_DDE}, the first term represents Markovian effects and the second convolutional integral term represents non-Markovian effects.
Applying the linear chain trick~\cite{hurtado2019generalizations}, when the kernel function $K(t)$ is represented by survival function of phase-type distribution $\mathrm{PH}(\mathbf{G},\boldsymbol{\alpha})$, whose precise definition is provided below, Eq.~\eqref{eq:simple_DDE} can be converted to the following form with the additional variables $\boldsymbol{\gamma}$:
\begin{align}
    \label{eq:simple_LCT}
    \frac{dx}{dt}&=ax+\sum_i \gamma_i,\\
    \frac{d\boldsymbol{\gamma}}{dt}&=\mathbf{G}\boldsymbol{\gamma}+x\boldsymbol{\alpha}.
\end{align}
That is, the convolutional integral of kernel functions in Eq.~\eqref{eq:simple_DDE} is replaced by a sum of additional variables $\boldsymbol{\gamma}$ in Eq.~\eqref{eq:simple_LCT}.
Then, to apply the quantum algorithms for solving linear differential equations, we use the linear chain trick to remove the integral terms in distributed delay differential equations.
\\
\indent The phase-type distribution $\mathrm{PH}(\mathbf{G},\boldsymbol{\alpha})$ is a mixture or convolution of exponential distributions~\cite{neuts1994matrix}, and its density function $f(t)$ and distribution function $F(t)$ are defined as follows:
\begin{align}
\label{eq:ft}
    f(t)&\equiv \boldsymbol{\alpha}^T \exp(t\mathbf{G})(-\mathbf{G}\mathbf{1}),\\
    \label{eq:Ft}
    F(t)&\equiv 1-\boldsymbol{\alpha}^T\exp(t\mathbf{G})\mathbf{1},
\end{align}
where $\mathbf{1}$ is a vector whose elements are all one, and $\mathbf{G}$ is a Hurwitz-Metzler matrix $G_{ii}<0$, $G_{ij}\geq0$ for $i\neq j$, and all eigenvalues of $\mathbf G$ are negative and $\boldsymbol{\alpha}$ is a probabilistic vector, that is, $\boldsymbol{\alpha}\geq 0$ and $\boldsymbol{\alpha}\cdot \mathbf{1}=1$. 
The matrix $\mathbf{G}$ satisfies $\sum_j G_{ij} \leq 0$ and at least one row sum is strictly negative.
The phase‑type distribution is the distribution of the absorption time of a finite-state Markovian process with at least one absorbing state.
For example, the Erlang distribution with parameters $(r,k)$ describes the distribution of dwelling time until $k$-th event in a Poisson process with rate $r$, whereas the hypoexponential distribution generalizes this distribution by allowing rate $r$ that is independent for each of events and time-dependent~\cite{neuts1994matrix,ibe2014fundamentals}.
The Coxian distribution further generalizes the hypoexponential distribution by allowing probabilistic absorption at each stage~\cite{cox1955use}.
Although these distributions correspond to a one-dimensional chain of Poisson processes, an arbitrary sequence of Poisson processes can be considered.
Moreover, any distribution supported on the nonnegative real line can be approximated arbitrarily closely by a phase-type distribution.

The probability that the state continues to enjoy the sequence of Markovian processes is defined as a survival function $S(t)$ as follows:
\begin{align}
    S(t)&\equiv \boldsymbol{\alpha}^T\exp(t\mathbf{G})\mathbf{1}\nonumber\\
    &=\int_0^t f(t) dt.
    \label{eq:St}
\end{align}
We introduce a variable $\Gamma$ to represent the convolution integral of the survival function $S(t)$ with the inflow $L(t)$. 
\begin{align}
    \Gamma(t)\equiv\int_0^t S(t-s)L(s)ds.
    \label{eq:Xt}
\end{align}
The integral $\Gamma(t)$ quantifies the accumulation of the inflow $L(t)$ under absorption (or outflow) dynamics dictated by a continuous-time Markovian process whose absorption time is distributed according to the phase-type distribution $\mathrm{PH}(\mathbf{G},\boldsymbol{\alpha})$.
The generalized linear chain trick can decompose $\Gamma(t)$ into $g$-dimensional additional variables $\boldsymbol{\gamma}(t)=(\gamma_1,\gamma_2,...,\gamma_g)$ defined as follows:
\begin{align}
    \boldsymbol{\gamma}(t)=\int_0^t\exp[(t-s)\mathbf{G}^T]\boldsymbol{\alpha}  L(s)ds,
    \label{eq:xt}
\end{align}
The time derivatives of these variables are derived as follows:
\begin{align}
    \frac{d\boldsymbol{\gamma}(t)}{dt}&=\mathbf{G}^T\int_0^t \exp[(t-s)\mathbf{G}^T]\boldsymbol{\alpha} f(s)ds+\boldsymbol{\alpha} L(t)\nonumber\\
    &=\mathbf{G}^T \boldsymbol{\gamma}(t)+\boldsymbol{\alpha} L(t),
    \label{eq:dxdt}
\end{align}
where the Leibniz formula is used in the transformation from the l.h.s. to the r.h.s. of the first line and the definition~\eqref{eq:xt} is used in the transformation from the first to the second line.
Furthermore, from Eqs.~\eqref{eq:St}--\eqref{eq:xt} $\Gamma(t)$ can be decomposed as follows:
\begin{align}
    \Gamma(t)&=\int_0^t L(s)\mathbf{1}^T\left(\exp[(t-s)\mathbf{G}^T]\boldsymbol{\alpha}\right)ds\nonumber\\
    &=\mathbf{1}^T\int_0^t \exp[(t-s)\mathbf{G}^T]\boldsymbol{\alpha} L(s)ds\nonumber\\
    &=\mathbf{1}^T\boldsymbol{\gamma}(t).
\end{align}
Therefore, the convolution $\Gamma(t)$ can be decomposed into a sum of the auxiliary variables $\boldsymbol{\gamma}(t)$.
From the definitions [Eqs.~\eqref{eq:Xt} and~\eqref{eq:xt}], the initial conditions are $\Gamma(t)=0$ and $\boldsymbol{\gamma}(t)=0$, as long as the integration is taken from $t=0$.
Note that although Ref~\cite{hurtado2019generalizations} assumes $\Gamma(t)$ and $L(t)\geq 0$ in the context of continuous Markovian processes, Eq.~\eqref{eq:GLCT_ODE}  can be mathematically obtained regardless of the sign, as is evident from the derivation. 

Let us exemplify a concrete application of the generalized linear chain trick when the kernel function is given by the Erlang distribution with parameters $(r,k)$, the resulting ordinary differential equation~\eqref{eq:GLCT_ODE} corresponds to that of usual linear chain trick as follows:
\begin{align}
    \frac{d\gamma_1}{dt}&=L(t)-r\gamma_1,\\
    \frac{d\gamma_i}{dt}&=r\gamma_{i-1}-r\gamma_{i} \quad i=2,3,\cdots,k. 
\end{align}
Therefore, the additional variables obey the linear differential equation composed of the inflow $L(t)$ and local interactions between $\boldsymbol{\gamma}$.
Each variable $\gamma_i$ represents the state in which $i-1$ events have occurred. At every time $t$, an inflow of $L(t)$ enters the initial state $\gamma_1$, where no events have yet taken place. 
With probability $r$, an event occurs and the process transitions to the subsequent state $\gamma_{i+1}$. 
Absorption at $\gamma_k$ signifies the termination of the process.

\subsection{Schr\"{o}dingerization technique}\label{subsec:Schrodingerization}
Schr\"{o}dingerization is a general framework of a quantum algorithm to solve linear differential equations via efficient Hamiltonian simulation~\cite{jin2024quantum,jin2024bquantum,jin2025schrodingerization} by embedding them in the Schr\"{o}dinger equation.
This method employs the so-called warped phase transformation to construct an augmented system by introducing one additional dimension to the original system of linear differential equations. 
In the frequency domain, the time evolution of this augmented system can be described by the Schr\"{o}dinger equation, enabling its efficient quantum simulation. 
Consequently, the solution of the original equations can be recovered from the time-evolved state of the augmented system via an inverse Fourier transform.

In the following, we introduce the procedure of the Schr\"{o}dingerization method according to Ref.~\cite{jin2024quantum}. 
Let us consider the following linear differential equation of $N$-dimensional vector $\mathbf{x}$
\begin{align}
    \frac{d\mathbf{x}}{dt}=\mathbf{Ax},
    \label{eq:linearODE}
\end{align}
where  the linear operator $\mathbf{A}$ is not anti-Hermitian in general.
The operator $\mathbf{A}$ can be decomposed as $\mathbf{A}=\mathbf{H}_1+i\mathbf{H}_2$  where the Hermitian part is $\mathbf{H}_1=(\mathbf{A}+\mathbf{A}^{\dagger})/2$ and the anti-Hermitian part is $i\mathbf{H}_2=-(\mathbf{A}-\mathbf{A}^{\dagger})/2$.
In the Schr\"{o}dingerization method, the Hermitian operator $H_1$ is assumed to be negative semi-definite; thus the operator $\mathbf{A}$ is semi-stable and the solution $\mathbf{x}(0)$ does not diverge.
Using the warped phase transformation $\mathbf{v}(t,p)= e^{-p}\mathbf{x}(t)$ for $p\geq 0$ , while we reverse the sign of $p$ for $p<0$ ; that is, the augmented variable $\mathbf{v}$ can be defined as 
\begin{align}
\label{eq:w}
    \mathbf{v}(t,p)\equiv e^{-|p|}\mathbf{x}(t).
\end{align}
Then, the state $\mathbf{x}(t)$ can be recovered from $\mathbf{v(t)}$ as $\mathbf{x(t)}=\int_0^{\infty}e^{p}\mathbf{v}(t)dp$.
By introducing the definition of $\mathbf{v}$ [Eq.~\eqref{eq:w}] into Eq.~\eqref{eq:linearODE}, we obtain the following partial differential equation
\begin{align}
    \frac{\partial\mathbf{v}}{\partial t}=-\mathbf{H}_1 \frac{\partial \mathbf{v}}{\partial p}-i\mathbf{H}_2 \mathbf{v}.
    \label{eq:pwpt1}
\end{align}
Moreover, we define the Fourier transform of $\mathbf{v}$ in $p$ with Fourier mode $\eta$ by $\tilde{\mathbf{v}}(t,\eta)\equiv\int_{-\infty}^{\infty}\exp(-i\eta p)\mathbf{v}(t,p)dp$. 
In terms of this representation, $\tilde{\mathbf{v}}$ obeys the following equation
\begin{align}
    \frac{\partial  \tilde{\mathbf{v}}}{\partial t}=-i(\eta \mathbf{H}_1+\mathbf{H}_2)\tilde{\mathbf{v}}.
    \label{eq:ptildewpt}
\end{align}
The partial derivative with respect to $p$ in time domain [Eq.~\eqref{eq:pwpt1}]  corresponds to imaginary multiplication in frequency domain [Eq~\eqref{eq:ptildewpt}], then Eq.~\eqref{eq:ptildewpt} is form of the Schr\"odinger-like equation.

In order to derive the Schr\"{o}dinger equation of quantum state in the Hilbert space from Eq.~\eqref{eq:ptildewpt}, we define the following quantum state 
\begin{align}
    \ket{\tilde{\mathbf{v}}}\equiv(\|\tilde{\mathbf{v}}\|)^{-1}\int_{-\infty}^{\infty}\tilde{\mathbf{v}}(t,\eta)\ket{\eta}d\eta.
    \label{eq:int_tilde}
\end{align}
By introducing this state vector into Eq.~\eqref{eq:ptildewpt}, its time evolution is described as follows: 
\begin{align}
        \label{eq:tilde_sim}
    \frac{\partial \ket{\tilde{\mathbf{v}}}}{\partial t}&=-i\mathbf{H}\ket{\tilde{\mathbf{v}}},\\
    \label{eq:sim_H}
    \mathbf{H}&\equiv\mathbf{H}_1\otimes\hat{\mathbf{\eta}} + \mathbf{H}_2\otimes \mathbf{I},
\end{align}
where $\mathbf{H}$ is Hermitian $\mathbf{H}=\mathbf{H}^{\dagger}$ and $\hat{\mathbf{\eta}}$ is a diagonal operator $\hat{\boldsymbol{\eta}}\ket{\eta}=\eta\ket{\eta}$.
To derive a form of Eq.~\eqref{eq:tilde_sim} simulatable on a quantum computer, we define the discretized version of the state $\ket{\tilde{\mathbf{v}}}$ as follows:
\begin{align}
    \ket{\tilde{\mathbf{v}}(t)}=\frac{1}{\|\tilde{\mathbf{v}}(t)\|}\sum_{k=0}^{N_{p}}\tilde{\mathbf{v}}(t,\eta_k)\ket{k},
    \label{eq:dis_tilde}
\end{align}
where the grid point of $p\in[-l/2,l/2]$ is denoted by $p_k\equiv kl/N_{p}-l/2$ , the number of grid points by $N_{p}$ , and the truncation interval by $l$ satisfying $e^{l/2}\simeq 0$. 
The discretized state $\ket{\tilde{\mathbf{v}}}$ evolves according to the Schr\"{o}dinger equation~\eqref{eq:tilde_sim} with the discretized Hamiltonian $\mathbf{H}$ represented as follows: 
\begin{align}
    \mathbf{H}&=\mathbf{H}_1\otimes \mathbf{D}+\mathbf{H}_2\otimes \mathbf{I},\\
    \mathbf{D}&=\mathrm{diag}(\boldsymbol{\mu}),
\end{align}
where $\mathbf{D}$ is a diagonalized matrix and $\boldsymbol{\mu}$ is a real vector $\mu_k=\frac{2\pi}{l}\left(k-\frac{N_{p}}{2}\right)$.
Since the state $\ket{\tilde{\mathbf{v}}}$ is obtained by applying the warped phase transformation and Fourier transformation, $\ket{\tilde{\mathbf{v}}(t)}=(\mathbf{I}\otimes \mathcal{F}_{p})\ket{\mathbf{x}_0}\sum_{j=N_{p}/2}^{N_{p}}\exp(-|p_k|)\ket{k}$.
Consequentially, by applying the inverse Fourier transformation to the state $\ket{\tilde{\mathbf{v}}}$, we can obtain the evolved state $\ket{\mathbf{v}(t)}=(\mathbf{I}\otimes \mathcal{F}_{p}^{-1})\ket{\tilde{\mathbf{v}}(t)}$ in the time domain.
Since the solution state $\ket{\mathbf{x}(t)}$ is embedded in the regime $p>0$ of the state $\ket{\mathbf{v}(t)}$, this state can be recovered by projecting $\ket{\mathbf{v}(t)}$ on $\mathbf{I}\otimes \sum_{k=N_{p}/2}^{N_{p}}\ket{k}{\bra{k}}$. 
The success probability is $\Theta(\|\mathbf{x}(t)\|^2/\|\mathbf{x}(0)\|^2)$ since the unitary transformation from $\ket{\mathbf{v}(t)}$ to $\ket{\mathbf{x}(0)}$ does not change the norm.
Quantum amplitude amplification is a more efficient approach to retrieve the solution state $\ket{\mathbf{x}(t)}$, yielding a quadratic speedup, reducing the complexity to $\mathcal{O}(\|\mathbf{x}(0)\|/\|\mathbf{x}(t)\|)$~\cite{brassard2000quantum}.
As a result, the total complexity of whole process of the Schr\"{o}dingerization is a product of those of the Hamiltonian simulation in frequency domain and that of readout of the solution state $\ket{\mathbf{x}(t)}$ from $\ket{\mathbf{v}(t)}$.
Let the sparsity be defined as $s=\max (s(\mathbf{H}_1),s(\mathbf{H}_2))$  and $\|\mathbf{H}\|_{\max}=\max\left(\|\mathbf{H}_1\|_{\max}/\epsilon,\,\|\mathbf{H}_2\|_{\max}\right)$.
Here, $s(\cdot)$ denotes the sparsity of the argument, and $\epsilon$ represents the allowable error.
According to Ref.~\cite{low2017optimal}, the query complexity of a sparse Hamiltonian simulation is $\mathcal{O}(s\,t\,\|\mathbf{H}\|_{\max}+\log\epsilon^{-1}/\log\log\epsilon^{-1})$.
Consequently, the total query complexity of the Schr\"{o}dingerization is
$\mathcal{O}\!\left((s\,t\,\|\mathbf{H}\|_{\max}+\log\epsilon^{-1}/\log\log\epsilon^{-1})\|\mathbf{x}(0)\|/\|\mathbf{x}(t)\|\right)$.\\
The overall gate complexity is the query complexity multiplied by $\mathcal{O}(m\log m + \log N)$, where $m$ is the number of precision bits.

\section{Quantum algorithm for solving linear distributed-delay differential equation}
In this section, we propose a quantum algorithm to efficiently solve linear distributed-delay differential equations and identify its applicable condition.
Since our algorithm relies on the Schr\"{o}dingerization method, the linear differential equations obtained obtained via the linear chain trick must be semi-stable.
Provided that the kernel function is characterized by phase-type distributions, we show that this requirement is satisfied if and only if the linear distributed-delay differential equation is also semi-stable.

\subsection{Quantum algorithm for solving linear distributed-delay differential equation}\label{subsec:our_algorithm}
In the following, we consider the following distributed delay differential equations or integro-differential equations
\begin{align}
\label{eq:DDE}
    \frac{dx_i}{dt}=\sum_{j=1}^N A_{ij}x_j(t) +\sum_{j=1}^N B_{ij}\int_0^t K_{ij}(t-s)x_j(s)ds,
\end{align}
where $\mathbf{x}=(x_1,x_2,...,x_N)^T$ is the $N$-dimensional state vector, $\mathbf{A}$ is the coefficient matrix of Markovian coupling, $\mathbf{B}$ is that of non-Markovian coupling, and $\mathbf{K}$ is the matrix composed of the phase-type distributed kernel functions.
We define the set $\mathcal{E}=\{(i,j)\mid B_{ij}\neq 0\}$; for each $(i,j)\in\mathcal{E}$, $K_{ij}$ is survival function [Eq.~\eqref{eq:St}] of the phase-type distribution $\mathrm{PH}(\mathbf{G}^{(i,j)},\boldsymbol{\alpha}^{(i,j)})$ defined by Eqs.~\eqref{eq:ft} and \eqref{eq:Ft}, otherwise $K_{ij}=0$.
The kernel function $K_{ij}$ for $(i,j)\in\mathcal{E}$ is normalized as $\int_0^\infty K_{ij}(t)dt=1$.
The sparsities of the matrices $\mathbf{A}$ and $\mathbf{B}$ are denoted by $s_A$ and $s_B$, respectively.
The maximum value of the dimension of $\boldsymbol{\alpha}^{(i,j)}$ is denoted by $g=\max_{(i,j)\in\mathcal E}\mathrm{dim}\,\boldsymbol{\alpha}^{(i,j)}$.
In this paper, we assume $s_B,g\geq1$, because Eq.~\eqref{eq:DDE}  is only a linear ordinary differential equation when $\mathbf{B}=\mathbf{0}$.
By applying the generalized linear chain trick [Eq.~\eqref{eq:GLCT_ODE}] to Eq.~\eqref{eq:DDE}, we obtain linear ordinary differential equations of augmented systems as follows:
\begin{align}
    \label{eq:GLCT_ODE}
    \frac{dx_i}{dt}&=\sum_{j=1}^N A_{ij}x_j+\sum_{j=1}^N B_{ij}\sum_{k=1}^{g^{(i,j)}} \gamma^{(i,j)}_{k},\\
    \label{eq:GLCT_ODE2}
    \frac{d\boldsymbol{\gamma}^{(i,j)}_{k}}{dt}&=\boldsymbol{\alpha}^{(i,j)}x_j+{\mathbf{G}^{(i,j)}}^T \boldsymbol{\gamma}^{(i,j)},\\
    \boldsymbol{\gamma}^{(i,j)}(t)&\equiv\int_0^t\exp\left[(t-s)\mathbf{G}^{(i,j)}\right]\boldsymbol{\alpha}^{(i,j)}  x_j(s)ds\nonumber\\
    &\hspace{3.5cm} \mathrm{for\,all\,}(i,j)\mathrm{\,in\,}\mathcal{E},
\end{align}
where for each $(i,j)\in\mathcal{E}$, $\boldsymbol{\gamma}^{(i,j)}$ is defined as an auxiliary vector of length $g^{(i,j)}=\dim \boldsymbol{\alpha}^{(i,j)}$.
Introducing the concatenated vector $\mathbf{y}$，Eqs.~\eqref{eq:GLCT_ODE} and \eqref{eq:GLCT_ODE2} can be combined into the following single linear differential equation
\begin{align}
\label{eq:augmented_ODE}
    \frac{d\mathbf{y}}{dt}&=\mathbf{C}\mathbf{y},\\
    \mathbf{y}&\equiv\begin{bmatrix}
\mathbf{x}\\\boldsymbol{\gamma}^{(1,d^{(1)}_1)}\\\boldsymbol{\gamma}^{(1,d^{(1)}_2)}\\\vdots\\\boldsymbol{\gamma}^{(N,d^{(N)}_{s_N})}
    \end{bmatrix},\quad
    \mathbf{C}\equiv\begin{bmatrix}
        \mathbf{A}&\mathbf{B}'\\
        \mathbf{E}&\mathbf{G}'
    \end{bmatrix},
\end{align}
and the augmented matrix $\mathbf{C}$ is defined by the following matrices and vectors
\begin{align}
\label{eq:B'}
    \mathbf{B}'&\equiv\begin{bmatrix}
        \mathbf{b}^{(1)}&\mathbf{O}&...&\mathbf{O}\\
        \mathbf{O}&\mathbf{b}^{(2)}&...&\mathbf{O}\\
        \vdots&\vdots&&\vdots\\
       \mathbf{O}&\mathbf{O}&...&\mathbf{b}^{(N)}
    \end{bmatrix},\\
    \mathbf{b}^{(i)}&\equiv\begin{bmatrix}
B_{i,d^{(i)}_1}\mathbf{1}^T_{g^{(i,d^{(i)}_1)}}&B_{i,d^{(i)}_2}\mathbf{1}^T_{g^{(i,d^{(i)}_2)}}&...&B_{i,d^{(i)}_{s_{i}}}\mathbf{1}^T_{g^{(i,d^{(i)}_{s_{i}})}}
    \end{bmatrix},\\
    \mathbf{E}&\equiv\begin{bmatrix}
        \boldsymbol{\alpha}^{(1,d^{(1)}_1)}\otimes (\mathbf{e}^{d^{(1)}_1})^T\\
        \boldsymbol{\alpha}^{(1,d^{(1)}_2)}\otimes (\mathbf{e}^{d^{(1)}_2})^T\\
        \vdots\\
        \boldsymbol{\alpha}^{(N,d^{(N)}_{s_N})}\otimes (\mathbf{e}^{d^{(N)}_{s_N}})^T
    \end{bmatrix},
    \quad\mathbf{e}^d_k=
\begin{cases}
1 & \text{if } k = d, \\
0 & \text{otherwise,}
\end{cases}\\
\label{eq:G'}
    \mathbf{G}'&\equiv\begin{bmatrix}
        \mathbf{G}^{(1,d^{(1)}_1)}&\mathbf{O}&...&\mathbf{O}\\
        \mathbf{O}&\mathbf{G}^{(1,d^{(1)}_2)}&...&\mathbf{O}\\
        \vdots&\vdots&&\vdots\\
        \mathbf{O}&\mathbf{O}&...&\mathbf{G}^{(N,d^{(N)}_{s_N})}
    \end{bmatrix},
\end{align}
where $\mathbf{1}_k$ denotes the all-ones vector of length $k$, $\mathbf{e}^{k}$ the $k$-th standard basis vector in $\mathbb{R}^N$, $s_i$ the number of nonzero entries in the $i$-th row of $\mathbf{B}$, $d^{(i)}_k$ the index of the $k$-th nonzero element of  the $i$-th row of $\mathbf{B}$, and the indices are ordered lexicographically with respect to $(i,j)$. 
Then, the length of the concatenated vector $\mathbf{y}$ is $N+\sum_{i=1}^N\sum_{j=1}^{s_i}g^{(i,j)}$.
The Schr\"{o}dingerization method requires the linear operator $\mathbf{C}$ to be semi-stable, ensuring that $\|\mathbf{y}(t)\|$ does not diverge.
Moreover, since the lengths of $\boldsymbol{\gamma}^{(i,j)}$ are not uniform, Eq.~\eqref{eq:augmented_ODE} is not suitable for an efficient simulation via the Schr\"{o}dingerization method, as one must be able to efficiently compute which entry of $\mathbf{y}$ corresponds to each $\boldsymbol{\gamma}^{(i,j)}_k$.

For efficient quantum simulation, the padded vector $\bar{\mathbf{y}}\in\mathbb{C}^{N(1+gs)}$ is introduced and Eq.~\eqref{eq:augmented_ODE} can be rewritten as follows (see Appendix~\ref{ap:zero-pad} for details of the zero-padded versions of the matrices and vectors [Eqs~\eqref{eq:B'}--\eqref{eq:G'}]): 
\begin{align}
\label{eq:padded_ODE}
    \frac{d\bar{\mathbf{y}}}{dt}&=\bar{\mathbf{C}}\bar{\mathbf{y}},\\
     \bar{\mathbf{y}}&\equiv\begin{bmatrix}
\mathbf{x}\\\bar{\boldsymbol{\gamma}}^{(1)}\\\bar{\boldsymbol{\gamma}}^{(2)}\\\vdots\\\bar{\boldsymbol{\gamma}}^{(N)}
\end{bmatrix},
\quad \bar{\boldsymbol{\gamma}}^{(i)}\equiv\begin{bmatrix}  \bar{\boldsymbol{\gamma}}^{(i,d^{(i)}_1)}\\\bar{\boldsymbol{\gamma}}^{(i,d^{(i)}_2)}\\\vdots\\\bar{\boldsymbol{\gamma}}^{(i,d^{(i)}_{s_i})}\\\mathbf{0}_{g(s-s_i)}
\end{bmatrix},
\quad \bar{\boldsymbol{\gamma}}^{(i,j)}\equiv\begin{bmatrix}
    \boldsymbol{\gamma}^{(i,j)}\\
    \mathbf{0}_{g-g^{(i,j)}}
\end{bmatrix},
\end{align}
where the $k$-th entry of $\boldsymbol{\gamma}^{(i,j)}$ is stored in the $(N + (i-1)gs + (j-1)g + k)$-th entry of the concatenated vector $\bar{y}$.

For an efficient quantum simulation, an efficient preparation of the initial state of the augmented system $\ket{\mathbf y(0)}$ is required.
By the definition [Eq.~\eqref{eq:xt}], the initial condition is always $\mathbf{x}(0)=\mathbf{0}$ as long as the convolutional integral begins at $t=0$.
Then, the augmented vector $\ket{\mathbf{y}(0)}$ can be efficiently prepared when the state vector $\ket{\mathbf{x}(0)}$ of the original system can be efficiently prepared. 
Note that, in general, whether the quantum state $\ket{\mathbf{x}}$ obtained by amplitude encoding a vector $\mathbf{x}$ can be efficiently prepared depends strongly on the structure of $\mathbf{x}$. 
In this work, we therefore assume that only these vectors $\ket{\mathbf{x}(0)}$ that are efficiently preparable are considered.

We consider the query and gate complexities of the simulation of the distributed delay differential equation via the Schr\"{o}dingerization.
The operator $\bar{\mathbf{C}}=\mathbf{H}_1+i\mathbf{H}_2$ is decomposed into the two Hermitian matrices $\mathbf{H}_1=(\bar{\mathbf{C}}+\bar{\mathbf{C}}^{\dagger})/2$ and $\mathbf{H}_2=(\bar{\mathbf{C}}-\bar{\mathbf{C}}^{\dagger})/(2i).$
From r.h.s. of Eq.~\eqref{eq:GLCT_ODE}, the time derivative of $\mathbf{x}$ is represented by a linear combination of at most $s_A$ Markovian coupling terms and $gs_B$ non-Markovian terms.
From r.h.s. of Eq.~\eqref{eq:GLCT_ODE2}, the derivative of $\boldsymbol{\gamma}$ is described by a linear combination of at most $g+1$ monomial terms.
Therefore, the sparsity of $\bar{\mathbf{C}}$ is $\mathcal{O}(s_A+gs_B)$.
Since the sparsity and norms of the matrices and vectors in the augmented system are preserved under zero-padding, we describe them in their pre-padded form.
Next, we consider the max norm of $\bar{\mathbf{C}}$. From the first identity in Eq.~\eqref{eq:GLCT_ODE}, the coefficients appearing in the differential equation for $\mathbf{x}$ are bounded by $\max(\|\mathbf{A}\|_{\max},\,\|\mathbf{B}\|_{\max})$, while from the second identity the coefficients on the right-hand side of the differential equation for $\boldsymbol{\gamma}$ are bounded by $\max(\|\mathbf{G}\|_{\max},\,1)$ since $\boldsymbol{\alpha}^{(i,j)}$ is a probability distribution. 
Consequently, $\|\mathbf{C}\|_{\max}=\max\big(1,\,\|\mathbf{A}\|_{\max},\,\|\mathbf{B}\|_{\max},\,\|\mathbf{G}\|_{\max}\big)$. 
For $\epsilon \ll 1$, even in the worst case we have $\|\mathbf{H}\|_{\max} \le \|\mathbf{C}\|_{\max}/\epsilon$.
Therefore, the whole query complexity for obtaining the solution vector $\ket{\mathbf{y}(t)}$ of Eq.~\eqref{eq:augmented_ODE} is at worst
\begin{align}
    \mathcal{O}\left(\left(\frac{1}{\epsilon}(s_A+gs_B)t\|\mathbf{C}\|_{\max}+\frac{\log\epsilon^{-1}}{\log\log\epsilon^{-1}}\right)\frac{\|\mathbf{y}(0)\|}{ \|\mathbf{y}(t)\|}\right).
\end{align}
Let us consider the complexity for obtaining the solution state $\ket{\mathbf{x}(t)}$ of original distributed delay differential equations.
As long as the domain of convolutional integral terms begin at $t=0$, the initial condition is always $\mathbf{x}(0)=\mathbf{0}$ by the definition [Eq.~\eqref{eq:xt}]; thus, we have $\|\mathbf{x}(0)\|=\|\mathbf{y}(0)\|$.
By performing amplitude amplification on $\ket{\mathbf{w}(t)}$ with respect to the basis of $\ket{\mathbf{x}(t)}$ restricted to components with $p>0$, one can recover the solution state $\ket{\mathbf{x}(t)}$.
Then, the query complexity for obtaining the solution state $\ket{\mathbf{x}(t)}$ is
\begin{align}
    \mathcal{O}\left(\left(\frac{1}{\epsilon}(s_A+gs_B)t\|\mathbf{C}\|_{\max}+\frac{\log\epsilon^{-1}}{\log\log\epsilon^{-1}}\right)\frac{\|\mathbf{x}(0)\|}{ \|\mathbf{x}(t)\|}\right).
\end{align}
An additional number of primitive quantum gates, scaling as the query complexity times $\mathcal{O}(m+\log(N(1+gs)))$, is required.
These complexities take the same form as those reported in~\cite{jin2024quantum} for solving linear ordinary or partial differential equations, except for $g$, despite the fact that our approach addresses distributed delay differential equations.
Note that the norm ratio $\|\|\mathbf x(0)\|\|/\|\|\mathbf x (t)\|\|$, or related quantities, are typical factors that arise in the complexity bounds of quantum algorithms for differential equations~\cite{berry2017quantum,krovi2023improved,fang2023time,jin2024quantum}.

\subsection{Semi-stability requirement}\label{subsec:semistability}
Since the Schr\"{o}dingerization method assumes that the linear ordinary differential equations are semi-stable, the ordinary differential equation derived by the linear chain trick [Eq.~\eqref{eq:augmented_ODE}] should be semi-stable.
Although some quantum algorithms~\cite{liu2021efficient,jin2024bquantum} require (semi-)stability of ordinary differential equations, it is preferable to state the applicable conditions in terms of the original distributed‑delay differential equations, since the problem is formulated in those forms.
Here we show that when the kernel functions are phase-type, the augmented linear ordinary differential equation [Eq.~\eqref{eq:augmented_ODE}] is semi-stable, if and only if the linear distributed delay differential equation [Eq.~\eqref{eq:DDE}] is semi-stable.
The stability of linear distributed delay differential equations is determined by the roots of the characteristic equation obtained via the Laplace transform, whereas that of linear ordinary differential equations is determined by the roots of the characteristic polynomial associated with the linear operator; for the class of problems considered in this paper, these two criteria coincide.

Consider the extended linear differential equation [Eq.~\eqref{eq:augmented_ODE}] whose characteristic matrix is
\begin{align}
\lambda \mathbf I_{N+M}-\mathbf{C}=
\begin{bmatrix} \lambda \mathbf I_N-\mathbf{A} & -\mathbf{B}'\\-\mathbf{E} & \lambda \mathbf I_M-\mathbf{G}' 
\end{bmatrix}. 
\end{align}
where $\mathbf{I}_k$ is the $k\times k$-size identity matrix and $M=\sum_{i,j}\dim \boldsymbol{\alpha}^{(i,j)}$. 
When $\mathbf{G}_{\lambda}\equiv\lambda \mathbf I_M-\mathbf{G}'$ is invertible $\det(\lambda \mathbf I_M-\mathbf{G}')\neq 0$, by using the Schur complement, the characteristic equation can be written as 
\begin{align} 
\det(\lambda \mathbf I_M-\mathbf{C}) &=\det(\mathbf{G}_{\lambda})\det(\lambda \mathbf I_M-\mathbf{A}-\mathbf{B}'\mathbf{G}_{\lambda}^{-1}\mathbf{E})\nonumber\\
&=0. 
\end{align}
According to the property of the phase-type distribution, the eigenvalues of every block matrix $\mathbf{G}^{(i,j)}$ are negative; consequently, all the eigenvalues of $\mathbf{G}'$ are negative.
Therefore, the linear differential equation of the extended system [Eq.~\eqref{eq:augmented_ODE}] is semi-stable precisely when every root $\lambda $ of the following characteristic equation satisfies $\mathrm{Re}(\lambda)\le 0$ and any root with $\mathrm{Re}(\lambda)=0$ is semi-simple:
\begin{equation} 
\det\big(\lambda \mathbf I_N-\mathbf{A}-\mathbf{B}'\mathbf{G}_{\lambda}^{-1}\mathbf{E}\big)=0.
\end{equation}
Next, we investigate the semi-stability condition of the linear distributed delay differential equation [Eq.~\eqref{eq:DDE}].
The Laplace transformation $\widehat{K}_{ij}(\lambda)$ of the kernel function $K_{ij}(t)$ which is defined by the survival function of the phase-type distribution $\mathrm{PH}(\mathbf G^{(i,j)},\boldsymbol{\alpha}^{(i,j)})$ is as follows:
\begin{equation} 
\widehat{K}_{ij}(\lambda)=\boldsymbol{\alpha}^{(i,j)}(\lambda \mathbf I_M-\mathbf{G}^{(i,j)})^{-1}\mathbf{1}. 
\end{equation} 
Noting that $\mathbf{G}'$ is block-diagonal with block matrices $\mathbf{G}^{(i,j)}$, the Laplace transform $\widehat{\mathbf{K}}(\lambda)$ of the matrix-valued kernel $\mathbf{K}(t)$ can be written as
\begin{align}
    \widehat{\mathbf K}(\lambda)=\mathbf{B}'\mathbf{G}_{\lambda}^{-1}\mathbf{E}.
\end{align}
Introducing $\widehat{\mathbf{K}}(\lambda)$, the Laplace transform of Eq.~\eqref{eq:DDE} yields
 \begin{align}
\lambda\widehat{\mathbf{x}}(\lambda)-\mathbf{x}(0)&=\mathbf{A}\,\widehat{\mathbf{x}}(\lambda)+\widehat{\mathbf{K}}(\lambda)\,\widehat{\mathbf{x}}(\lambda)\nonumber\\
&=\mathbf A \widehat{\mathbf x}(\lambda)+\mathbf{B}'\mathbf{G}_{\lambda}^{-1}\mathbf{E}\widehat{\mathbf x}(\lambda).
 \end{align}
Then, the characteristic equation is 
\begin{align} 
\det\big(\lambda \mathbf I_N-\mathbf{A}-\widehat{K}(\lambda)\big) &=\det(\lambda \mathbf I_N-\mathbf{A}-\mathbf{B}'\mathbf{G}_{\lambda}^{-1}\mathbf{E})\nonumber\\
&=0. 
\end{align}
Because the characteristic equation governing the augmented linear differential equation coincides with the characteristic equation of the distributed-delay differential equation, the semi-stability conditions are identical: all roots $\lambda$ must lie in the closed left half-plane and any root on the imaginary axis must be semi-simple. 
Consequently, when the distributed delay differential equation is semi-stable, the Schr\"{o}dingerization technique is applicable to the linear differential equation obtained through the generalized linear chain trick, and the proposed method applies to semi-stable distributed delay equations with kernel functions that are characterized by phase-type distributions.

\section{APPLICATION}
To showcase the efficacy of the proposed method, we exemplify the concrete models of linear distributed delay differential equations [Eq.~\eqref{eq:DDE}].
In particular, we demonstrate the applicability of our algorithm to both classical and quantum regimes by considering the generalized master equation and the Redfield equation.

\subsection{Generalized master equation}
The generalized master equation is the non-Markovian version of the master equation, which represents a general continuous Markovian process.
As a network model, this equation has been extensively applied to the study of diverse phenomena, ranging from random walks\cite{allegrini2003generalized} and animal movement~\cite{giuggioli2009generalized} to the reaction-transport systems~\cite{fedotov2002non}.
\begin{align}
\label{eq:GME}
    \frac{d\mathbf{p}(t)}{dt}=\int_0^t \mathbf{R}(t-\tau)\mathbf{p}(\tau)d\tau,
\end{align}
where $\mathbf{p}(t)$ denotes the probability distribution and $\mathbf{R}(t)$ the time-dependent generation matrix.
The generation matrix $\mathbf{R}$ is defined from the waiting time distribution , and the master equation of a Markovian process is recovered when $\mathbf{R}(t)=r_{ij}\delta(t)$ with the amplitude parameter $r_{ij}$ and the delta function $\delta(t)$.
Clearly, the generalized master equation corresponds to the special case of Eq.~\eqref{eq:DDE} where the Markovian coupling matrix $\mathbf{A}$ is absent.
Since the generation matrix $\mathbf{R}$ is typically integrable and decaying, the waiting-time distribution is often given exactly by a phase-type distribution, in which case the generalized linear chain trick can be applied.
The proposed algorithm yields the solution state $\ket{p(t)}$ with query complexity $\mathcal{O}(\|\mathbf p(0)\|/\|\mathbf p(t)\|(\epsilon^{-1}  g s_r t \|\mathbf{r}\|_{\max}+\log\epsilon^{-1}/\log\log{\epsilon}^{-1}))$, where the unnormalized kernel function $R_{ij}(t)=r_{ij}K_{ij}(t)$ is decomposed by the amplitude $r_{ij}$ and the normalized kernel function $\int_0^\infty K_{ij}(t)=1$, and $s_r$ is the sparsity of the matrix $r$.

\subsection{Modified Redfield equation}\quad
Th Redfield equation describes the non-Markovian dynamics of general open quantum systems, which can be derived from the Schr\"{o}dinger equation of a composite system of a principal system and an environment by removing the degree of the environment via the Nakajima-Zwanzig projection~\cite{nakajima1958quantum,zwanzig1960ensemble,breuer2002theory}.
Given that the Hamiltonian of the system interacting with the environment is defined by $H_{\mathrm{tot}}=H_{\mathrm{S}}+H_{\mathrm{SE}}+H_{\mathrm{E}}$, where $H_{\mathrm{S}}$ is the system Hamiltonian, $H_{\mathrm{SE}}$ is the interaction Hamiltonian, and $H_{\mathrm{E}}$ is the environment Hamiltonian, we assume that $H_S$ and $H_{\mathrm{SE}}$ commute as $[H_{\mathrm{S}},H_{\mathrm{SE}}]=0$.
Otherwise, the sparsities of corresponding linear operators are lost because they grow and spread in the Hilbert space during time evolution (see Appendix~\ref{ap:RF} for the details of the assumptions and derivation).
When the interaction Hamiltonian is decomposed into the sum of $h$ products of the system operator $T_m$ and the environment operator $V_m$ as $H_{\mathrm{SE}}=\sum_{m=1}^{h}T_m\otimes V_m$, the modified Redfield equation under the commutation assumption can be written as follows:
\begin{align}
\label{eq:Redfield}
\frac{d}{dt}\rho_{\mathrm S}(t)
&= -i[H_{\mathrm S},\rho_{\mathrm S}(t)]\nonumber\\
&-\sum_{m,n=1}^h \int_0^t d\tau\;
\Big(
W_{mn}(\tau)\,[T_m,\,T_n\,\rho_{\mathrm S}(t-\tau)]\nonumber\\
&+ W_{nm}^*(\tau)\,[T_m,\,\rho_{\mathrm S}(t-\tau)\,T_n]
\Big),
\end{align}
where $\rho_S$ denotes the density operator of the principal system and $W_{mn}(t)\equiv\mathrm{Tr}[V_m(t)V_n\rho_{\mathrm{E}}]$ the correlation function of the environment with respect to $V_m$ and $V_n$.
This commuting case of a quantum system, not limited to the Redfield equation, is known as the dephasing model, and has been extensively investigated for mathematical convenience and experimental significance~\cite{yu2003qubit,marche2025exceptional}.
Furthermore, when the spectral density of the environment is Ohmic or Lorentz, the correlation function $W_{mn}(t)$ is given by an exponential decay or its linear combination; thus, the assumption that the waiting time distribution is phase-type is naturally satisfied.
The above matrix-form equation can be converted into the vectorized form in the Fock-Liouville space; thus, the solution state $\ket{\rho_{\mathrm S}(t)}$ can be efficiently computed (see Appendix~\ref{ap:FL} for details).
Note that $\|\rho_{\mathcal S}(t)\|_{\mathrm{HS}}$ is non-increasing in the dephasing model because when the density operator is expanded in the basis of the engenvectors of the Hamiltonian $H_{\mathrm{S}}$, the dynamics reduces to a decay of the coherence.\\

\section{Conclusion}
In this paper, we develop a quantum algorithm to efficiently solve the linear distributed delay differential equations by embedding them in ordinary differential equations via the generalized linear chain trick.
Through the Schr\"{o}dingerization framework, the solution state of the ordinary differential equation can be computed with an efficient Hamiltnonian simulation algorithm.
The requirements of the proposed method is as follows: the semi-stability of the distributed delay differential equations, the sparse coupling among variables, and the kernel functions characterized by a phase-type distribution.
Moreover, the initial state of the augmented system can be efficiently prepared when that of the original system is efficiently preparable since the initial conditions of the auxiliary variables are zero.
We demonstrate the application of the proposed method to the two cases: the generalized Master equation and the Redfield equation of the dephasing model.
Our results extend the recent development of the quantum algorithms for solving differential equations into the regime of non-Markovian dynamics. 

Nevertheless, there are mainly two limitations of our proposed algorithm: the linearity of memory effects and the dependence on the structure of the phase-type distribution.
To embed in the Schr\"{o}dinger equation, our method handles only the linear cases of distributed delay differential equations.
Because recent papers~\cite{liu2021efficient,tanaka2023polynomial} provide efficient solutions for nonlinear differential equations on quantum computers, our method might also be extended to the cases of nonlinear memory effects.
Our algorithm assumes that the kernel functions can be represented by the survival function of the phase-type distribution, however, this type of distributions can approximate arbitrary non-negative distributions within an arbitrary allowable error.
Then, in an approximate setting, the proposed method may be generalized to the cases with accommodate kernel functions governed by arbitrary probability distributions. 
\begin{acknowledgments}
This work is by MEXT Quantum Leap Flagship Program (MEXT Q-LEAP) Grant No. JPMXS0120319794, JST COI-NEXT Grant No. JPMJPF2014, and JST CREST JPMJCR24I3.
WS is supported by JST ACT-X, Japan, Grant No. JPMJAX25CF.
\end{acknowledgments}

\appendix
\section{Zero-padded matrices and vectors}\label{ap:zero-pad}
For an efficient quantum simulation, the matrices and vectors in Eq.~\eqref{eq:augmented_ODE} are resized by zero-padding as follows: 
\begin{align}
\label{eq:ap:padded_ODE}
    \frac{d\bar{\mathbf{y}}}{dt}&=\bar{\mathbf{C}}\bar{\mathbf{y}},\\
     \bar{\mathbf{y}}&\equiv\begin{bmatrix}
\mathbf{x}\\\bar{\boldsymbol{\gamma}}^{(1)}\\\bar{\boldsymbol{\gamma}}^{(2)}\\\vdots\\\bar{\boldsymbol{\gamma}}^{(N)}
\end{bmatrix},
\quad \bar{\boldsymbol{\gamma}}^{(i)}\equiv\begin{bmatrix}  \bar{\boldsymbol{\gamma}}^{(i,d^{(i)}_1)}\\\bar{\boldsymbol{\gamma}}^{(i,d^{(i)}_2)}\\\vdots\\\bar{\boldsymbol{\gamma}}^{(i,d^{(i)}_{s_i})}\\\mathbf{0}_{g(s-s_i)}
\end{bmatrix},
\quad \bar{\boldsymbol{\gamma}}^{(i,j)}\equiv\begin{bmatrix}
    \boldsymbol{\gamma}^{(i,j)}\\
    \mathbf{0}_{g-g^{(i,j)}}
\end{bmatrix},\\
\bar{C}&\equiv\begin{bmatrix}
    A&\bar{B}'\\
    \bar{S}&\bar{G}'
\end{bmatrix},
\end{align}
where
\begin{widetext}
\begin{align}
\label{eq:padB'}
\bar{\mathbf{B}}'&\equiv\begin{bmatrix}
        \bar{\mathbf{b}}^{(1)}&\mathbf{O}&...&\mathbf{O}\\
        \mathbf{O}&\bar{\mathbf{b}}^{(2)}&...&\mathbf{O}\\
        \vdots&\vdots&&\vdots\\
       \mathbf{O}&\mathbf{O}&...&\bar{\mathbf{b}}^{(N)}
    \end{bmatrix},&&\\
\bar{\mathbf{b}}^{(i)}&\equiv\begin{bmatrix}
\bar{\mathbf{b}}^{(i,d^{(i)}_1)}&\bar{\mathbf{b}}^{(i,d^{(i)}_2)}&...&\bar{\mathbf{b}}^{(i,d^{(i)}_{s_i})}&\mathbf{0}^T_{g(s-s_i)}
    \end{bmatrix},\quad  
&\bar{\mathbf{b}}^{(i,j)}&\equiv\begin{bmatrix}
        B_{i,d^{(i)}_1}\mathbf{1}^T_{g^{(i,d^{(i)}_1)}}&\mathbf{0}^T_{g-g^{(i,d^{(i)}_1)}}
    \end{bmatrix},\\
\bar{\mathbf{E}}&\equiv\begin{bmatrix}
        \boldsymbol{\alpha}^{(1,d^{(1)}_1)}\otimes (\mathbf{e}^{d^{(1)}_1})^T\\
        \mathbf{0}_{g-g^{(1,d^{(1)}_1)}}\otimes \mathbf{0}_{N}\\
        \boldsymbol{\alpha}^{(1,d^{(1)}_2)}\otimes (\mathbf{e}^{d^{(1)}_2})^T\\
        \mathbf{0}_{g-g^{(1,d^{(1)}_2)}}\otimes \mathbf{0}_{N}\\
        \vdots\\
        \boldsymbol{\alpha}^{(N,d^{(N)}_{S_N})}\otimes (\mathbf{e}^{d^{(N)}_{s_N}})^T\\
        \mathbf{0}_{g-g^{(N,d^{(N)}_{s_N})}}\otimes \mathbf{0}_{N}
        \end{bmatrix},
    \quad&\mathbf{e}^{d}_{k}&=
\begin{cases}
1 & \text{if } k = d, \\
0 & \text{otherwise,}
\end{cases}\\
\label{eq:padG'}
    \bar{\mathbf{G}}'&\equiv\begin{bmatrix}
        \bar{\mathbf{G}}^{(1)}&\mathbf{O}&...&\mathbf{O}\\
        \mathbf{O}&\bar{\mathbf{G}}^{(2)}&...&\mathbf{O}\\
        \vdots&\vdots&\vdots&&\vdots\\
        \mathbf{O}&\mathbf{O}&...&\bar{\mathbf{G}}^{(N)}
    \end{bmatrix},&&\\
\bar{\mathbf{G}}^{(i)}&\equiv\begin{bmatrix}
        \bar{\mathbf{G}}^{(i,d^{(i)}_1)}&\mathbf{O}&...&\mathbf{O}&\mathbf{O}\\
        \mathbf{O}&\bar{\mathbf{G}}^{(i,d^{(i)}_2)}&...&\mathbf{O}&\mathbf{O}\\
        \vdots&\vdots&&\vdots&\vdots\\
        \mathbf{O}&\mathbf{O}&...&\bar{\mathbf{G}}^{(i,d^{(i)}_{s_i})}&\mathbf{O}\\
        \mathbf{O}&\mathbf{O}&\cdots&\mathbf{O}&\mathbf{O}_{g(s-s_i)\times g(s-s_i)}
    \end{bmatrix},
&\bar{\mathbf{G}}^{(i,j)}&\equiv\begin{bmatrix}
        \mathbf{G}^{(i,j)}&\mathbf{O}_{g^{(i,j)},g-g^{(i,j)}}\\
        \mathbf{O}_{g-g^{(i,j)},g^{(i,j)}}&\mathbf{O}_{g-g^{(i,j)},g-g^{(i,j)}}
    \end{bmatrix}.
\end{align}
\end{widetext}
The dimensions of the matrices and vectors are as follows: $\bar{\mathbf{B}}'\in\mathbb{C}^{N\times sN}$, $\bar{\mathbf{b}}^{(i)}\in\mathbb{C}^{gs}$, $\bar{\mathbf{E}}\in\mathbb{C}^{gN\times N}$, $\bar{\mathbf{G}}'\in \mathbb{C}^{gsN\times gsN}$, $\bar{\mathbf{G}}^{(i)}\in \mathbb{C}^{gs\times gs}$, and $\bar{\mathbf{G}}^{(i,j)}\in \mathbb{C}^{g\times g}$.

\section{Redfield equation of dephasing model}\label{ap:RF}
Here, we derive the Redfield equation of the dephasing model[Eq.~\eqref{eq:Redfield}].
As explained in the main text, the interaction Hamiltonian $H_{\mathrm{SE}}$ can be written as a sum of $h$ products:
\begin{align}
H_{\mathrm{SE}} = \sum_{k=1}^h T_k \otimes V_k.
\end{align}
By introducing the interaction picture, the operator $X$ at time $t$ can be written as $\tilde{X}(t)\equiv  e^{iH_0t}Xe^{-iH_0t}$ with the reference Hamiltonian $H_0=H_{\mathrm S}+H_{\mathrm E}$, and 
\begin{align}
\tilde H_{\mathrm{SB}}(t) &= \sum_{k=1}^h \tilde T_k(t)\otimes \tilde V_k(t), \\
\tilde T_k(t)& = e^{i H_{\mathrm S} t} T_k e^{-i H_{\mathrm S} t},\\
\tilde V_k(t) &= e^{i H_{\mathrm E} t} V_k e^{-i H_{\mathrm E} t}.
\end{align}
Under the Born approximation, the density operator can be taken as $\rho=\rho_{\mathrm{S}}\otimes\rho_{\mathrm{E}}$, and the system dynamics in the interaction picture can be written as follows:  
\begin{align}
\label{eq:ap:int_rhoS}
\frac{d}{dt}\tilde \rho_{\mathrm S}(t)
= -\int_0^t d\tau\;
\mathrm{Tr}_{\mathrm E}\Big(
[\tilde H_{\mathrm{SE}}(t),\,[\tilde H_{\mathrm{SE}}(\tau),\,
\tilde \rho_{\mathrm S}(t-\tau)\otimes \rho_{\mathrm E}]]
\Big).
\end{align}
By substituting $H_{\mathrm{SE}}=\sum_m^h T_m\otimes V_m$ into Eq.~\eqref{eq:ap:int_rhoS}, we obtain the Redfield equation as follows:
\begin{align}
\label{eq:ap:RF}
\frac{d}{dt}\tilde \rho_{\mathrm S}(t)
&= -\sum_{m, n=1}^h \int_0^t d\tau\;
\Big(
W_{mn}(\tau)\,[\tilde T_m(t),\,\tilde T_n(\tau)\,\tilde \rho_{\mathrm S}(t-\tau)]\nonumber\\
&+ W_{nm}(-\tau)\,[\tilde T_m(t),\,\tilde \rho_{\mathrm S}(t-\tau)\,\tilde T_n(\tau)]
\Big),
\end{align}
where
\begin{align}
W_{mn}(\tau) = \mathrm{Tr}_{\mathrm E}\big[\tilde V_m(\tau)\,V_n\,\rho_{\mathrm E}\big].
\end{align}
From the commuting assumption of the dephasing model such that $[H_{\mathrm{S}},T_m]=0$ for all $m$, it can be easily shown $\tilde{T}_m(t)=T_m$. 
Then, Eq.~\eqref{eq:ap:RF} can be simplified as follows:
\begin{align}
\label{eq:ap:RF_dp}
\frac{d}{dt}\tilde \rho_{\mathrm S}(t)
&= -\sum_{m,n=1}^h \int_0^t d\tau\;
\Big(
W_{mn}(\tau)\,[T_m,\,T_n\,\tilde \rho_{\mathrm S}(t-\tau)]\nonumber\\
&+ W_{nm}^*(\tau)\,[T_m,\,\tilde \rho_{\mathrm S}(t-\tau)\,T_n]
\Big).
\end{align}
In the Schr\"{o}dinger picture, Eq.~\eqref{eq:ap:RF_dp} in the interaction picture can be rewritten as follows:
\begin{align}
\frac{d}{dt}\rho_{\mathrm S}(t)
&= -i[H_{\mathrm S},\rho_{\mathrm S}(t)]\nonumber\\
&-\sum_{m,n=1}^h \int_0^t d\tau\;
\Big(
W_{mn}(\tau)\,[T_m,\,T_n\,\rho_{\mathrm S}(t-\tau)]\nonumber\\
&+ W_{nm}^*(\tau)\,[T_m,\,\rho_{\mathrm S}(t-\tau)\,T_n]
\Big).
\end{align}

\section{Fock-Liouville space}\label{ap:FL}
In the Fock-Liouville space, by vectorizing the density operator, its time evolution can be represented as linear differential equations, unlike in the Hilbert space~\cite{manzano2020short}.
In short, the mapping of the matrix $\mathbf{Y}\mapsto \mathbf{XYZ}$ in the Hilbert space can be transformed in the Fock-Liouville space as follows:

\begin{align}
    \ket{\mathbf{Y}_v}&\mapsto (\mathbf{Z}^T\otimes \mathbf{X})\ket{\mathbf{Y}_v},\\
    \ket{\mathbf{Y}_v}&\equiv\begin{bmatrix}
        Y_{11}\\Y_{21}\\\vdots\\Y_{N1}\\Y_{12}\\\vdots\\Y_{NN}
    \end{bmatrix},
\end{align}
where $\mathbf{X}$, $\mathbf{Y}$, and $\mathbf{Z}\in\mathbb{C}^{N\times N}$ and $\ket{\mathbf{Y}_v}\in \mathbb{C}^{N^2}$.
Furthermore, the inner product $\mathrm{Tr}[A^{\dagger}B]$ can be represented by $\braket{A_v|B_v}$ in the Fock-Liouville space.
Applying this conversion to Eq.~\eqref{eq:Redfield}, we obtain the vectorized form of the Redfield equation
\begin{align}
        \frac{d\ket{\rho_v(t)}}{dt}&=-i(I\otimes H_{\mathrm S}+H_{\mathrm S}^T\otimes I) \ket{\rho_v(t)}\nonumber\\
        &+\sum_{m, n=1}^h\int_0^t\left\{ W_{mn}(\tau)[I\otimes T_{m}T_{n}+T_m^T\otimes T_n]\right.\nonumber\\
        &\left.+W^*_{nm}(\tau)[T_n^T\otimes T_m+(T_nT_m)^T\otimes I]\right\}\ket{\rho_v(t-\tau)} d\tau.
\end{align}
Following the proposed method, we derive the linear differential equation for the augmented system [Eq.~\eqref{eq:augmented_ODE}]. 
Denoting its linear operator by $\mathbf{C}_{\mathrm{RF}}$, the maximum norm is given by $ \|\mathbf{C}_{\mathrm{RF}}\| = \left\{ 2\|H\|_{\max}, \; h^{2}\|T_{m}\|_{\max}^{2}, \; \max_{m,n}\|G^{(m,n)}\|_{\max} \right\}$ and the sparsity is expressed as $s(\mathbf{C}_{\mathrm{RF}}) = s(H) + 4h^{2}\max_{m} s(T_{m})^{2} + \max_{m,n,c} s(G_c^{(m,n)})$, where the real and imaginary parts of correlation function $W_{m,n}(t)$ are assumed to be the normalized survival functions of the phase-type distribution $\mathrm{PH}(\mathbf G_r^{(m,n)},\boldsymbol \alpha_r^{(m,n)})$ and $\mathrm{PH}(\mathbf G_i^{(m,n)},\boldsymbol \alpha_i^{(m,n)})$, respectively.

\end{document}